# Device Engineering and Performance Optimization of $Cu_2NiGeS_4$ Thin-Film Solar Cells with $In_2S_3/MoTe_2$ Charge-Selective Layers: A Computational Study

Md Tashfiq Bin Kashem [a,*], and Hasib Md Abid Bin Farid [a]

[a] Department of Electrical and Electronic Engineering, Ahsanullah University of Science and Technology, Dhaka – 1208, Bangladesh

[*] Corresponding author: E-mail address: tashfiq.eee@aust.edu

**Abstract:** The pursuit of efficient and sustainable thin-film photovoltaics increasingly demands absorber materials that combine strong optical absorption with earth-abundant and environmentally benign constituents. $Cu_2NiGeS_4$ (CNGS) has emerged as a promising quaternary chalcogenide absorber owing to its favorable optoelectronic properties and high absorption coefficient, yet its photovoltaic potential remains comparatively underexplored, particularly in conjunction with optimized charge-selective layers. Here, we introduce and comprehensively investigate an Al/AZO/$In_2S_3$/CNGS/$MoTe_2$/Au solar-cell architecture using SCAPS-1D, with $In_2S_3$ and $MoTe_2$ serving as the electron and hole-transport layers respectively. The analysis of energy-band profiles, electric fields, carrier distributions, and generation-recombination characteristics reveals the mechanisms governing carrier separation and extraction across the heterojunctions. Notably, $MoTe_2$ serves a dual function as the hole-transport layer and a secondary absorber, extending photon utilization toward longer wavelengths beyond the CNGS absorption edge. Systematic optimization of layer thickness, doping density, bulk and interface defect densities, recombination coefficients, parasitic resistances, temperature, and illumination intensity identifies the key factors limiting device performance. The optimized device is predicted to achieve a power conversion efficiency (PCE) of 28.44% with a open-circuit voltage ($V_{OC}$) of 0.984 V, short-circuit current density ($J_{SC}$) of 34.39 mA/cm$^2$, and fill factor (FF) of 84.04%, under AM1.5G illumination at 300 K. This simulated efficiency exceeds the previously reported 6.25-21.17% range for CNGS-based solar cells considered in this study. The findings establish $In_2S_3$/CNGS/$MoTe_2$ as a promising platform for next-generation thin-film photovoltaics and provide physically grounded design guidelines for absorber optimization, interface engineering, and future experimental realization.

**Keywords:** Thin-film solar cell; CNGS; $In_2S_3$; $MoTe_2$; SCAPS-1D

## 1. Introduction

The rapid growth in global energy demand and the urgent need to reduce greenhouse gas emissions have intensified the search for sustainable and economically viable photovoltaic (PV) technologies. Among the various PV technologies, thin-film solar cells have attracted considerable attention because of their low material consumption, compatibility with large-area and flexible substrates, and potential for low-cost fabrication [1]–[4]. In recent decades, remarkable progress has been achieved with thin-film absorbers such as CdTe and Cu(In,Ga)$Se_2$ (CIGS) [5], [6], whose laboratory-scale power conversion efficiencies now exceed 23% [7]. Nevertheless, the widespread commercialization of these technologies remains constrained by the scarcity of indium and tellurium, together with concerns regarding the toxicity of cadmium. These limitations have stimulated extensive research into earth-abundant, environmentally benign, and low-cost absorber materials capable of delivering comparable photovoltaic performance while ensuring long-term material sustainability [8].

Kesterite-based quaternary chalcogenides have consequently emerged as one of the most promising alternatives for next-generation thin-film photovoltaics. Materials such as $Cu_2ZnSnS_4$ (CZTS), $Cu_2ZnSnSe_4$ (CZTSe), and their mixed-anion alloys possess high optical absorption coefficients (>$10^4$ cm$^{-1}$), suitable direct bandgaps, and consist primarily of abundant, non-toxic elements [9], [10]. Despite these advantages, their device efficiencies remain substantially below the theoretical limit. This performance deficit is primarily attributed to severe Cu-Zn cation disorder, the formation of antisite defects and secondary-phases, and the associated band-tail states, all of which enhance non-radiative recombination and reduce carrier lifetime [11]–[13]. Consequently, compositional engineering through isovalent cation substitution has become an effective strategy for mitigating intrinsic defect formation while preserving the favorable optoelectronic characteristics of kesterite absorbers [14], [15].

Among the emerging quaternary chalcogenides, $Cu_2NiGeS_4$ (CNGS) has recently attracted increasing attention as a potential photovoltaic absorber [16]–[18]. Replacing Zn with Ni and Sn with Ge is expected to suppress the formation of detrimental antisite defects that commonly limit conventional kesterite absorbers, while simultaneously improving structural stability and defect tolerance. Experimental studies have demonstrated that CNGS can be synthesized as highly crystalline thin films using cost-effective deposition techniques [19], [20], exhibiting a direct bandgap close to 1.8 eV together with a large optical absorption coefficient exceeding $10^4$ $cm^{-1}$. First-principles calculations have further confirmed its favorable electronic structure, strong visible-light absorption, and promising photovoltaic characteristics, indicating that CNGS satisfies many of the fundamental requirements of an efficient thin-film absorber [18], [19], [21], [22].

Despite these encouraging material properties, research on CNGS-based solar cells remains at a very early stage. To date, only a limited number of experimental and theoretical investigations have been reported, most of which focus primarily on material synthesis, structural characterization, or the intrinsic electronic and optical properties of CNGS. A few numerical studies have explored the feasibility of CNGS absorbers using conventional device architectures [17], [18], [23], demonstrating its photovoltaic potential. However, systematic optimization of CNGS-based solar cells through appropriate transport-layer engineering remains largely unexplored. In particular, the combined use of diindium trisulfide ($In_2S_3$) as the electron transport layer (ETL) and molybdenum ditelluride ($MoTe_2$) as the hole transport layer (HTL) has not yet been comprehensively investigated for CNGS-based photovoltaic devices.

The selection of $In_2S_3$ and $MoTe_2$ is motivated by their favorable optoelectronic properties and band alignment with the CNGS absorber. $In_2S_3$ is a cadmium-free wide-bandgap semiconductor that has been extensively investigated as a high-transparency buffer layer because of its excellent optical transmission, suitable electron affinity, and low interfacial recombination characteristics [24]–[29]. In contrast, $MoTe_2$ possesses a relatively narrow bandgap of approximately 1.1 eV, high hole mobility, and appropriate valence-band alignment for efficient hole extraction [30]–[32]. Owing to its relatively small bandgap, $MoTe_2$ can additionally absorb near-infrared photons transmitted through the CNGS layer [33], [34], potentially contributing to photocurrent generation while simultaneously functioning as the hole transport layer. Such a dual functionality offers an attractive opportunity for improving light harvesting and carrier collection within a single-junction device.

Motivated by these considerations, this work presents a comprehensive numerical investigation of an Al/AZO/$In_2S_3$/CNGS/$MoTe_2$/Au thin-film solar cell using the one-dimensional solar cell capacitance simulator (SCAPS-1D) simulator. Unlike previous reports, this study systematically examines the coupled influence of absorber, ETL, and HTL properties together with bulk and interface defects, resistive losses, operating temperature, illumination intensity, intrinsic recombination mechanisms, and capacitance-voltage characteristics. Furthermore, the underlying device physics is explained through detailed analyses of the energy-band structure, electric-field distribution, carrier concentrations, generation and recombination profiles, depletion characteristics, and Mott–Schottky behavior.

The optimized device demonstrates a high theoretical power conversion efficiency, highlighting the significant potential of CNGS as a next-generation earth-abundant absorber material. More importantly, the present work proposes practical design guidelines for transport-layer engineering and defect management in CNGS-based thin-film photovoltaics, thereby providing valuable insights for future experimental realization and optimization of this promising photovoltaic technology.

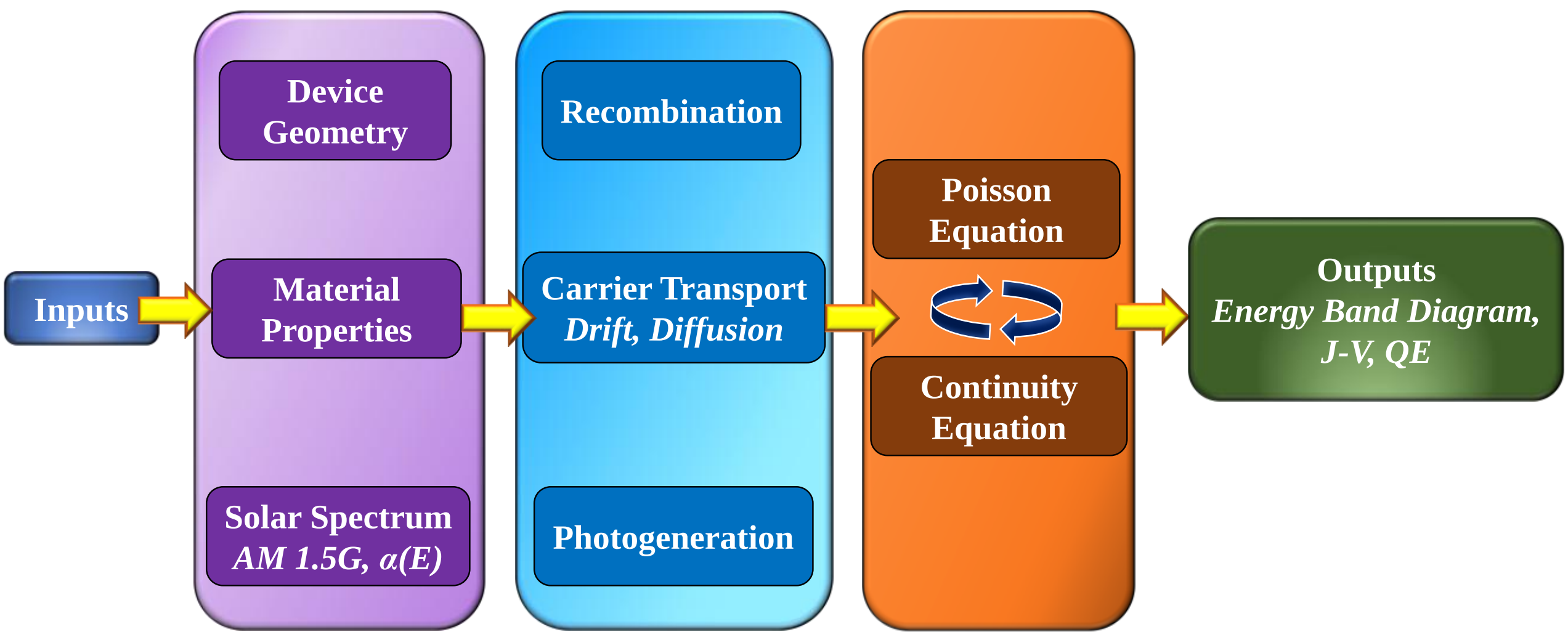

Fig. 1. Schematic workflow of the SCAPS-1D numerical simulation framework.

## 2. Simulation methodology

The numerical simulations presented in this work were performed using the SCAPS-1D simulator (version 3.3.12), which was developed at the Department of Electronics and Information Systems (ELIS), Ghent University, Belgium [35]–[37]. SCAPS-1D has been extensively employed for the design and optimization of thin-film photovoltaic devices and has consistently demonstrated excellent agreement with experimentally reported characteristics for a wide range of absorber materials and device architectures [38]–[45]. Because of its robust numerical framework and proven predictive capability, SCAPS-1D has become one of the most widely accepted simulation platforms for investigating the electrical and optical behavior of emerging photovoltaic technologies [46]–[50]. The overall simulation procedure adopted in this study is illustrated in Fig. 1.

The device characteristics are obtained by self-consistently solving the coupled nonlinear Poisson and carrier continuity equations under steady-state conditions using the drift-diffusion transport model, where SCAPS-1D employs a Newton-Raphson iterative scheme to achieve numerical convergence. The electrostatic potential distribution within the device is governed by Poisson's equation,

$$\frac{\partial^2 \Phi(x)}{\partial x^2} = -\frac{q}{\epsilon}\left[p(x) - n(x) + N_D^+ - N_A^- \pm N_{def}(x)\right] \quad (1)$$

where $\Phi$ denotes the electrostatic potential, q is the elementary charge, $\epsilon$ is the dielectric permittivity, n and p represent the electron and hole concentrations, respectively, $N_D^+$ and $N_A^-$ are the ionized donor and acceptor concentrations, and $N_{def}$ corresponds to the defect density.

The conservation of charge carriers is described by the electron and hole continuity equations,

$$\begin{cases} \text{Electron: } \frac{\partial n}{\partial t} = \frac{1}{q}\frac{\partial J_n}{\partial x} + (G_n - R_n) & (2) \\ \text{Hole: } \frac{\partial p}{\partial t} = -\frac{1}{q}\frac{\partial J_p}{\partial x} + (G_p - R_p) & (3) \end{cases}$$

where $J_n$ and $J_p$ denote the electron and hole current densities, while $G$ and $R$ represent the carrier generation and recombination rates, respectively.

Carrier transport within the device is described by the drift-diffusion equations,

$$\begin{cases} \text{Electron: } J_n = qn\mu_n\varepsilon + qD_n\frac{\partial n}{\partial x} & (4) \\ \text{Hole: } J_p = qp\mu_p\varepsilon - qD_p\frac{\partial p}{\partial x} & (5) \end{cases}$$

where $\mu$ is the carrier mobility, $\varepsilon$ is the electric field, and $D$ is the diffusion coefficient. The first term in each equation represents drift transport driven by the internal electric field, whereas the second term accounts for carrier diffusion resulting from concentration gradients. These coupled differential equations are solved iteratively until numerical convergence is achieved, yielding the spatial distributions of energy bands, electric field, carrier concentrations, generation and recombination rates, and current density throughout the device.

The optical generation profile was calculated using the wavelength-dependent absorption coefficients ($\alpha$) of the constituent semiconductor layers, which were directly imported into SCAPS-1D (Table 1). The absorption spectra of $In_2S_3$ and $MoTe_2$ were adopted from experimental reports [51], [52], whereas that of the CNGS absorber was obtained from a published first-principles (ab initio) investigation [18]. The simulator internally determines the spatial and wavelength dependent photogeneration profile by combining these wavelength-dependent absorption data with the incident photon flux, $N_{photon}$ [37],

$$G(x,\lambda) = \alpha(x,\lambda)N_{photon}(x,\lambda) \qquad (6)$$

Integration of the above equation over all wavelengths of the incoming photon flux provides the spatial generation rate,

$$G(x) = \int_{\lambda_{min}}^{\lambda_{max}} G(x,\lambda)\, d\lambda \qquad (7)$$

Carrier recombination within the device is modeled by considering Shockley-Read-Hall (SRH), radiative, and Auger recombination mechanisms. The total recombination rate is therefore expressed as [53],

$$R_{total} = R_{SRH} + R_{rad} + R_{Auger} \qquad (8)$$

The SRH recombination rate is given by,

$$R_{SRH} = \frac{np - n_i^2}{\tau_p(n + n_1) + \tau_n(p + p_1)} \tag{9}$$

where $n$ and $p$ are the electron and hole concentrations, $n_i$ is the intrinsic carrier concentration, $\tau_n$ and $\tau_p$ denote the electron and hole lifetimes, while $n_1$ and $p_1$ are the equilibrium carrier concentrations associated with the trap energy level.

Radiative recombination is described by,

$$R_{rad} = B(np - n_i^2) \tag{10}$$

where $B$ is the radiative recombination coefficient.

The Auger recombination rate is expressed as,

$$R_{Auger} = \left(C_n n + C_p p\right)(np - n_i^2) \tag{11}$$

where $C_n$ and $C_p$ are the Auger recombination coefficients for electrons and holes, respectively.

The principal photovoltaic parameters are extracted from the current density-voltage (J-V) characteristics [54],

$$J(V) = J_{Dark} - J_{Light} = J_0 \left[\exp\left(\frac{qV}{kT}\right) - 1\right] - J_{Light} = J_0 \left[\exp\left(\frac{qV}{kT}\right) - 1\right] - J_{SC} \tag{12}$$

Where $J$ is the current density, $J_{Dark}$ is the current density under dark, $J_{Light}$ is the current density under illumination, which is also the short circuit current density ($J_{SC}$), $J_0$ is the reverse saturation current density, $V$ is the terminal voltage, $k$ is the Boltzmann constant, and $T$ is the absolute temperature.

The open circuit voltage ($V_{OC}$) is calculated as,

$$V_{OC} = V(J = 0) = \frac{kT}{q} \ln\left[\frac{J_{SC}}{J_0} + 1\right] \tag{13}$$

while the fill factor and power conversion efficiency are calculated using,

$$FF = \frac{V_m J_m}{V_{OC} J_{SC}} \tag{14}$$

$$PCE = \frac{V_{OC} J_{SC} FF}{P_{in}} \tag{15}$$

where $V_m$ and $J_m$ are the voltage and current density at the maximum power point, respectively, and $P_{in}$ is the incident optical power density.

In addition to the conventional photovoltaic parameters, SCAPS-1D was employed to extract the spatial distributions of energy bands, electric field, carrier concentrations, generation rate, recombination rate, capacitance-voltage characteristics, depletion width, and apparent carrier concentration, thereby providing detailed insight into the internal operating physics of the proposed solar cell. Unless otherwise stated, all simulations were carried out under the standard AM1.5G solar spectrum with an incident power density of 1000 $Wm^{-2}$ at an operating temperature of 300 K.

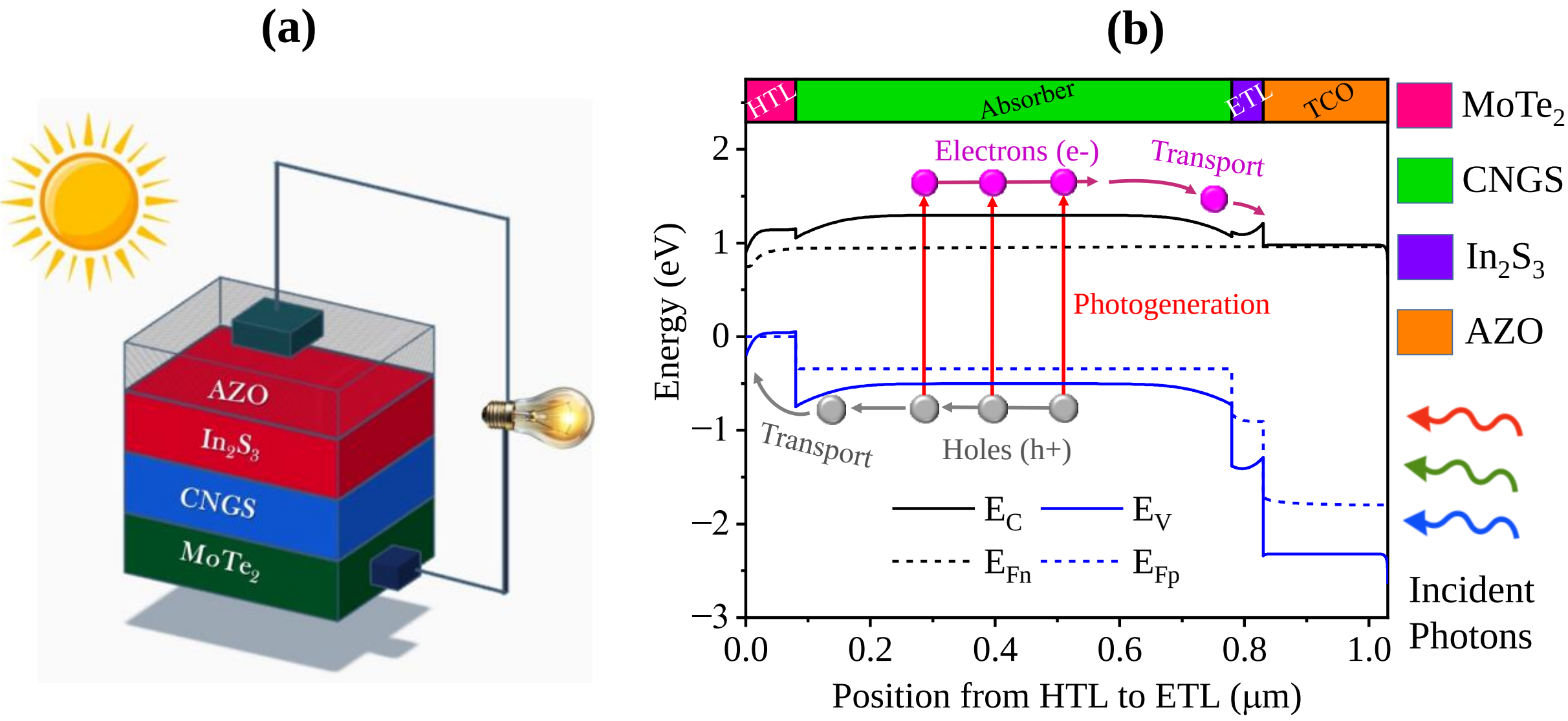

Fig. 2. (a) Schematic representation of the Al/AZO/$In_2S_3$/CNGS/$MoTe_2$/Au thin-film solar cell architecture employed in the simulations, and (b) corresponding energy-band diagram illustrating photogenerated electron-hole pairs and their transport pathways. CNGS serves as the primary photoactive absorber, while $In_2S_3$ and $MoTe_2$ function as the ETL and HTL, respectively, facilitating selective carrier extraction and suppressing transport of the opposite charge carriers.

## 3. Device structure and material parameters

The proposed solar cell architecture investigated in this work consists of the configuration Al/AZO/$In_2S_3$/CNGS/$MoTe_2$/Au, as schematically illustrated in Fig. 2(a) and the equilibrium energy-band diagram of the structure is presented in Fig. 2(b). Aluminum-doped zinc oxide (AZO) is employed as the transparent conducting oxide (TCO) owing to its high optical transparency in the visible region, excellent electrical conductivity, low cost, and environmental compatibility [55], [56]. The AZO layer facilitates efficient transmission of incident photons to the active region while providing a low-resistance pathway for electron collection. Aluminum (Al) and gold (Au), possessing work functions of 4.2 eV and 5.1 eV, respectively [57], are employed as the front and back metallic contacts to establish favorable energy-level alignment with the adjacent semiconductor layers and facilitate efficient charge extraction [58]. The front contact was maintained at ground potential, whereas the terminal voltage was imposed at the back contact. The resulting heterostructure promotes directional transport of photogenerated electrons toward the AZO/$In_2S_3$ side and holes toward the $MoTe_2$/Au side, thereby limiting carrier accumulation and associated recombination losses.

The material parameters used in the simulations, including bandgap, electron affinity, dielectric constant, carrier mobility, effective density of states, and defect characteristics, were adopted from previously reported experimental and theoretical studies and are summarized in Tables 1 and 2.

Table 1. Input parameters of the materials: band gap ($E_g$), electron affinity ($\chi$), relative dielectric permittivity ($\varepsilon_r$), effective density of states in the conduction band ($N_c$), effective density of states in the valence band ($N_v$), electron mobility ($\mu_n$), hole mobility ($\mu_h$), shallow uniform acceptor density ($N_A$), shallow uniform donor density ($N_D$), and defect density ($N_t$)

| Parameters | AZO | $In_2S_3$ | CNGS | $MoTe_2$ |
|---|---|---|---|---|
| Thickness (μm) | 0.2 | 0.05 | 0.7 | 0.08 |
| $E_g$ (eV) | 3.3 | 2.5 | 1.8 | 1.1 |
| $\chi$ (eV) | 4.5 | 4.25 | 4.3 | 4.2 |
| $\varepsilon_r$ | 9 | 7.72 | 6.56 | 13 |
| $N_c$ (1/cm$^3$) | $2.2 \times 10^{18}$ | $2.2 \times 10^{18}$ | $2.58 \times 10^{18}$ | $10^{15}$ |
| $N_v$ (1/cm$^3$) | $1.8 \times 10^{19}$ | $1.8 \times 10^{19}$ | $4.54 \times 10^{18}$ | $10^{17}$ |
| $\mu_n$ (cm$^2$/Vs) | 100 | 100 | 100 | 110 |
| $\mu_h$ (cm$^2$/Vs) | 25 | 25 | 25 | 426 |
| $N_D$ (1/cm$^3$) | $10^{18}$ | $10^{17}$ | 0 | 0 |
| $N_A$ (1/cm$^3$) | 0 | 0 | $10^{16}$ | $5 \times 10^{17}$ |
| *Defect type* | Neutral | Neutral | Neutral | Neutral |
| $N_t$ (1/cm$^3$) | $10^{14}$ | $10^{14}$ | $10^{14}$ | $10^{14}$ |
| Reference | [18] | [25] | [17] | [49] |
| $\alpha$ (1/cm) | | | | |

Table 2. Input parameters for interface and bulk defects

| Parameters | $In_2S_3$/CNGS interface | CNGS/$MoTe_2$ interface | Bulk $NaSnCl_3$ |
|---|---|---|---|
| Defect type | Neutral | Neutral | Neutral |
| Electron capture cross section (cm$^2$) | $10^{-19}$ | $10^{-19}$ | $10^{-15}$ |
| Hole capture cross section (cm$^2$) | $10^{-19}$ | $10^{-19}$ | $10^{-15}$ |
| Energetic distribution | Single | Single | Single |

| Reference for defect energy level | Above $E_i$ | Above $E_i$ | Above $E_i$ |
|---|---|---|---|
| Energy with respect to reference (eV) | 0.0 | 0.0 | 0.0 |
| Total density ($cm^{-2}$) | $10^{10}$ | $10^{10}$ | $10^{14}$ $cm^{-3}$ |

## 4. Results and discussion

### 4.1. Effect of thickness and doping density of absorber, ETL and HTL

The performance of the solar cell is governed by a strongly coupled interplay between the absorber and charge-selective transport layers (ETL and HTL). The systematic variation of thickness and doping density of CNGS, $In_2S_3$, and $MoTe_2$ in this work reveals that device operation is primarily controlled by the balance among (i) built-in electric field distribution, (ii) carrier transport resistance, and (iii) recombination dynamics at bulk and interface regions.

Within the CNGS absorber, the device response is determined by a trade-off between enhanced electric-field strength and increased recombination with rising acceptor density (Fig. 3). Higher absorber doping strengthens band bending at both heterojunction interfaces, improving carrier separation and increasing $V_{OC}$ (Fig. 3a) [59], [60]. However, this benefit is counteracted by increased recombination and reduced diffusion length, which suppresses $J_{SC}$ (Fig. 3b). The resulting non-monotonic behavior of FF and PCE reflects the competition between improved internal field formation and deteriorated carrier lifetime (Fig. 3c-d). In contrast, increasing absorber thickness enhances photon absorption and generation of electron–hole pairs, leading to an overall improvement in current collection and efficiency, provided that recombination remains within tolerable limits.

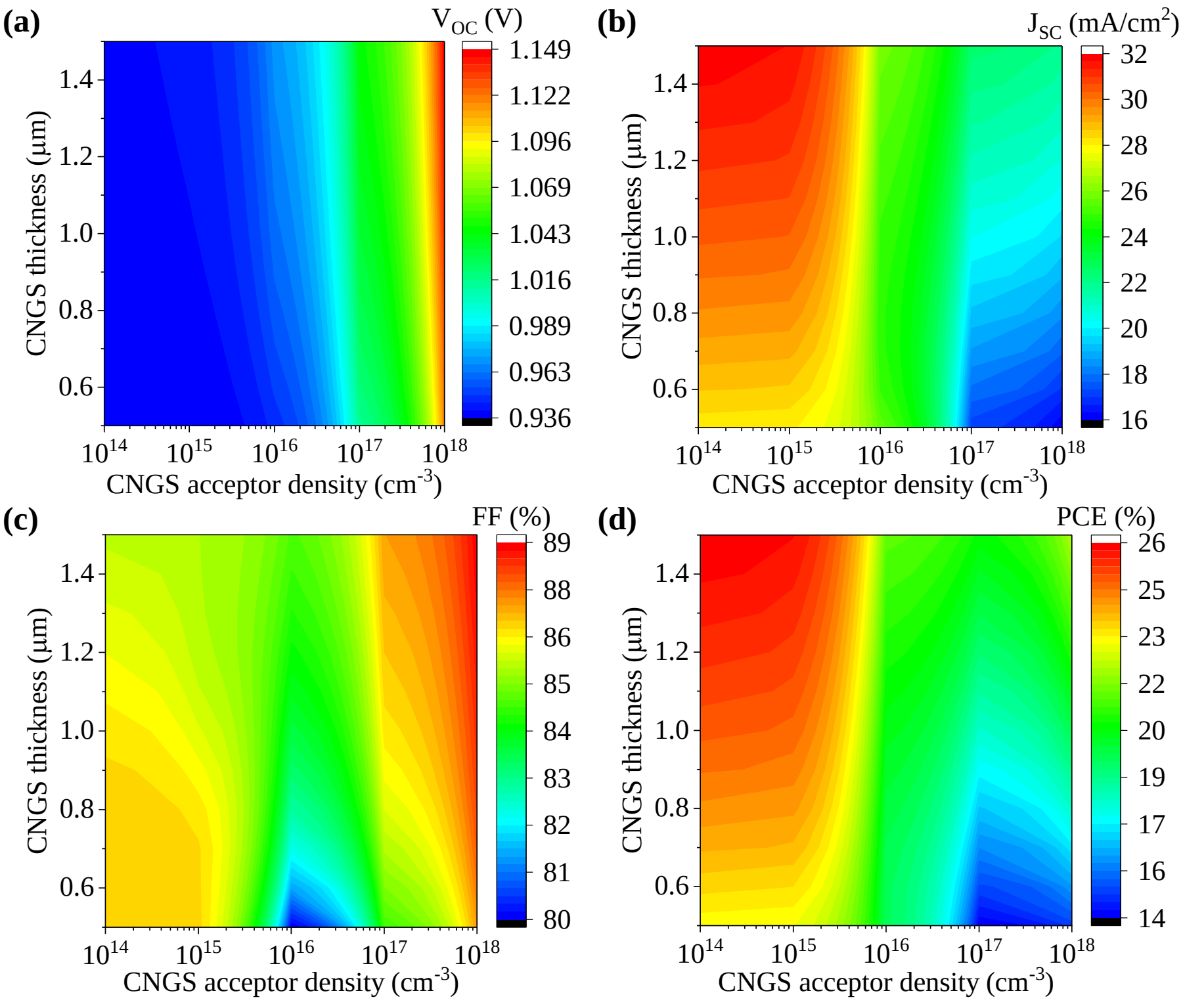

Fig. 3. Two-dimensional maps showing the simultaneous effects of CNGS absorber thickness and acceptor density on the photovoltaic performance of the device: (a) $V_{OC}$, (b) $J_{SC}$, (c) FF, and (d) PCE. All other material parameters were kept fixed at their default values, as listed in Tables 1 and 2.

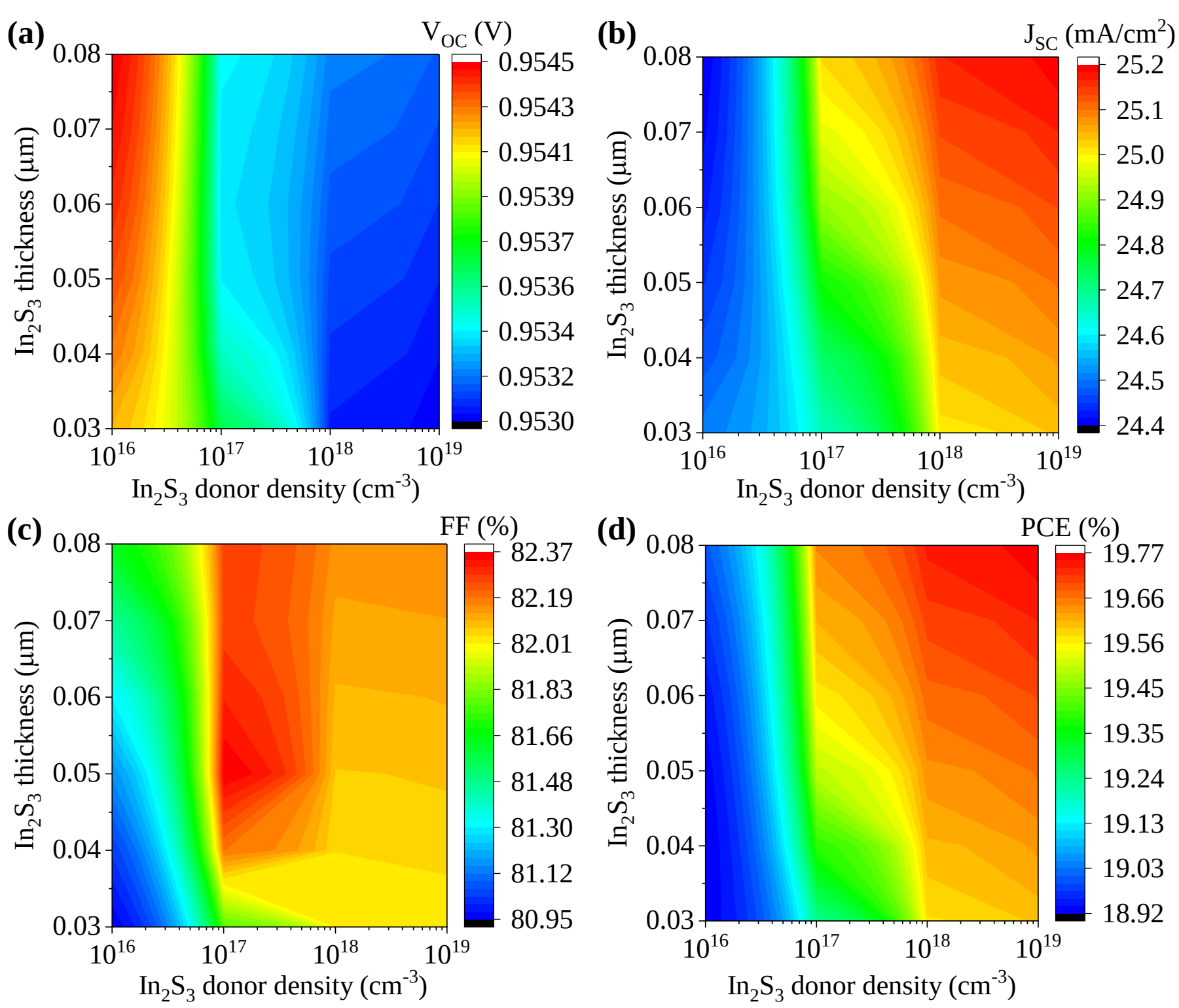

Fig. 4. Two-dimensional maps showing the simultaneous effects of $In_2S_3$ ETL thickness and donor density on the photovoltaic performance of the device: (a) $V_{OC}$, (b) $J_{SC}$, (c) FF, and (d) PCE. All other material parameters were kept fixed at their default values, as listed in Tables 1 and 2.

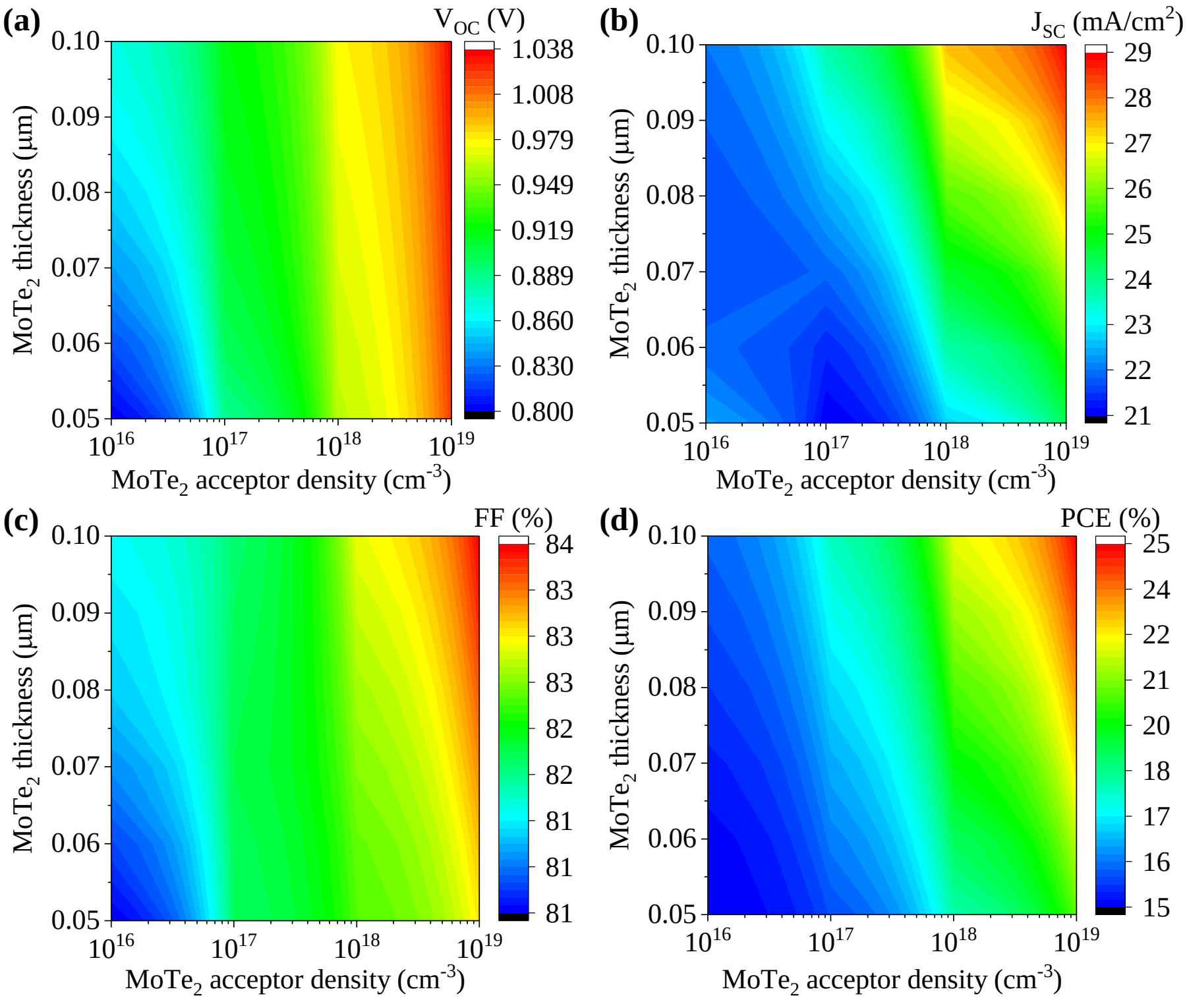

Fig. 5. Two-dimensional maps showing the simultaneous effects of $MoTe_2$ HTL thickness and acceptor density on the photovoltaic performance of the device: (a) $V_{OC}$, (b) $J_{SC}$, (c) FF, and (d) PCE. All other material parameters were kept fixed at their default values, as listed in Tables 1 and 2.

The $In_2S_3$ ETL plays a distinct role dominated by electron transport and front-interface selectivity (Fig. 4). Increasing donor density improves electron conductivity and facilitates carrier extraction, leading to higher $J_{SC}$ and FF (Fig. 4b-c) [61], [62]. However, excessive donor concentration modifies the electrostatic potential distribution at the $In_2S_3$/CNGS interface, leading to saturation and slight degradation in FF and a modest reduction in $V_{OC}$ (Fig. 4a). Similarly, increasing ETL thickness enhances electron selectivity and suppresses minority-carrier leakage toward the front contact, thereby improving all photovoltaic parameters within the investigated range. This indicates that the ETL operates primarily in a regime where improved selectivity outweighs transport penalties.

The $MoTe_2$ HTL exhibits the most monotonic behavior (Fig. 5). Increasing acceptor density improves hole extraction efficiency by enhancing conductivity and strengthening band bending at the rear junction [63]. This reduces back-contact recombination and improves charge collection efficiency. Likewise, increasing HTL thickness improves hole selectivity and suppresses interfacial recombination losses at the metal contact, resulting in simultaneous enhancement of $V_{OC}$, $J_{SC}$, FF, and PCE (Fig. 5a-d) [64]. The increase in $J_{SC}$ with $MoTe_2$ thickness is also supported by its additional photoabsorbing contribution identified in Section 4.1, whereby longer-wavelength photons transmitted through CNGS are harvested within the lower-bandgap $MoTe_2$ layer. Thus, within the investigated thickness range, the combined benefits of hole-selective transport and additional photon harvesting outweigh any transport penalty associated with increasing $MoTe_2$ thickness.

### 4.2. CNGS bulk defect density

The influence of bulk defect density in the CNGS absorber was investigated to evaluate the defect tolerance of the proposed device architecture and to assess the impact of non-radiative recombination centers on photovoltaic performance (Fig. 6). In quaternary chalcogenide semiconductors such as CNGS, bulk defects can originate from deviations in stoichiometry during film growth, incomplete crystallization, secondary-phase formation, and intrinsic point defects including vacancies, antisite defects, and interstitials [12], [65]. These imperfections introduce localized energy states within the bandgap that act as carrier trapping and recombination centers. Consequently, controlling the defect density is critical for achieving high-efficiency devices, as excessive defects can severely limit carrier lifetime, diffusion length, and collection efficiency. In the present simulations, the CNGS bulk defects were represented by neutral single-level defects located at the mid-bandgap, with equal electron and hole capture cross sections of $10^{-15}$ cm$^2$. Mid-bandgap states can serve as particularly effective SRH recombination centers because their energetic position facilitates sequential capture of electrons and holes. Accordingly, increasing the density of these neutral defects primarily influences carrier trapping, lifetime, and SRH recombination.

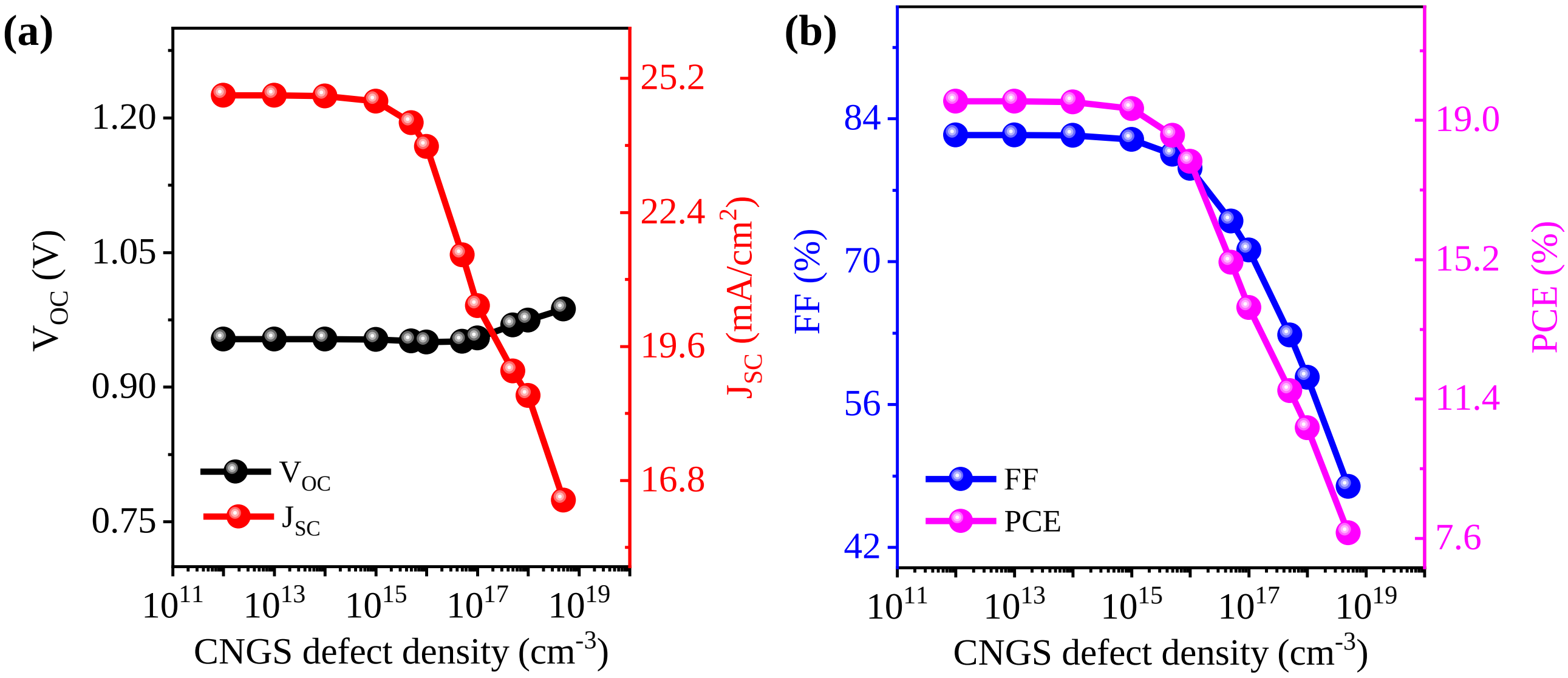

Fig. 6. Effect of bulk defect density of the absorber, CNGS on (a) $V_{OC}$ and $J_{SC}$, and (b) FF, and PCE of the device. All other material parameters were kept fixed at their default values, as listed in Tables 1 and 2.

As observed from Fig. 6a, $V_{OC}$ remains nearly constant up to a bulk defect density of approximately $10^{17}$ cm$^{-3}$, indicating that the recombination activity associated with defects is initially insufficient to significantly alter the attainable quasi-Fermi level splitting. Beyond this threshold, a gradual increase in $V_{OC}$ is observed. Although this trend appears counterintuitive from the conventional expectation of increasing SRH losses with defect density, non-monotonic and anomalous $V_{OC}$ responses associated with bulk defects have also been identified in numerical studies of thin-film heterojunction solar cells [66], while increases in $V_{OC}$ with absorber defect density have been reported for certain absorber systems in SCAPS-1D simulations [67]. Under open-circuit conditions, $V_{OC}$ is governed by the steady-state electron and hole populations and their corresponding quasi-Fermi-level splitting, which are obtained self-consistently with defect occupation and recombination. Because the defects considered here are electrically neutral, the slight increase in $V_{OC}$ should not be attributed to defect-induced enhancement of the built-in electric field or junction electrostatics. Rather, it is more appropriately regarded as a modest model-dependent response arising from the self-consistent interplay among carrier populations, defect occupation, and recombination under open-circuit conditions. Importantly, this small increase occurs concurrently with deterioration in $J_{SC}$, FF, and PCE and therefore does not imply that increasing the CNGS bulk defect density is beneficial to device operation.

In contrast, $J_{SC}$, FF, and PCE remain nearly unchanged only up to a defect density of approximately $10^{15}$ cm$^{-3}$, after which all three parameters gradually deteriorate (Fig. 6a-b). This degradation is primarily associated with enhanced SRH recombination through defect-assisted trapping processes, consistent with the strong sensitivity of thin-film chalcogenide solar cells to elevated absorber defect densities reported in recent numerical studies [63], [68]. The SRH carrier lifetime is inversely proportional to defect density and can be expressed as

$$\tau_{n,p} = \frac{1}{\sigma_{n,p}\, N_t\, v_{th}} \qquad (16)$$

where $\sigma$ is the carrier capture cross section, $v_{th}$ is the thermal velocity, and $N_t$ is the defect density. As the defect concentration increases, the carrier lifetime decreases, leading to a reduction in the minority-carrier diffusion length $\left(L_{n,p} = \sqrt{\frac{kT}{q}\, \mu_{n,p}\, \tau_{n,p}}\right)$, where $\mu$ is the carrier mobility. The resulting reduction in carrier lifetime and diffusion length lowers the probability that photogenerated carriers reach the charge-selective interfaces before recombination. Consequently, a larger fraction of carriers recombines within the CNGS absorber, resulting in a progressive reduction in current collection efficiency and a corresponding decline in $J_{SC}$.

The reduction in FF with increasing defect density originates from the same recombination-driven mechanism. Enhanced defect-assisted recombination increases carrier losses throughout the absorber and modifies the illuminated J-V characteristics, thereby reducing the maximum extractable electrical power. Since PCE is governed by the combined evolution of $V_{OC}$, $J_{SC}$, and FF, the efficiency follows the dominant trends of $J_{SC}$ and FF. Although a slight increase in $V_{OC}$ is observed at high defect densities, the accompanying reduction in carrier lifetime and collection efficiency has a much stronger influence on device operation, leading to a net decline in PCE.

## 4.3. Interface defect density

The quality of the heterointerfaces plays a decisive role in determining the performance of thin-film solar cells, as interfaces represent the primary regions where photogenerated carriers are separated and extracted. In practical devices, interface defects may arise from lattice mismatch, interfacial disorder, dangling bonds, compositional interdiffusion, incomplete surface passivation, and the formation of secondary phases during layer deposition [69], [70]. These defects introduce localized energy states within the bandgap that act as SRH recombination centers, thereby reducing carrier lifetime and collection efficiency [71]. Consequently, evaluating the influence of interface defect density is essential for assessing device stability, fabrication tolerance, and the feasibility of experimental realization.

The effect of defect density at the $In_2S_3$/CNGS interface reveals a relatively high degree of defect tolerance (Fig. 7). $V_{OC}$, $J_{SC}$, FF, and PCE remain nearly unchanged up to an interface defect density of approximately $10^{16}$ $cm^{-2}$, indicating that recombination losses at the front junction remain limited within this range. At low interface defect densities, the built-in electric field efficiently separates photogenerated carriers before they can be captured by interfacial trap states. Once the defect density exceeds $10^{16}$ $cm^{-2}$, all photovoltaic parameters gradually decrease. The increasing density of interfacial traps enhances SRH recombination at the $In_2S_3$/CNGS junction, reducing the minority-carrier lifetime and increasing carrier losses during the extraction process. As a result, the quasi-Fermi level splitting decreases, leading to a reduction in $V_{OC}$, while the diminished carrier collection efficiency lowers $J_{SC}$. The combined impact of these losses results in a gradual degradation of FF and overall device efficiency.

In contrast, the CNGS/$MoTe_2$ interface exhibits significantly greater sensitivity to defect formation. $V_{OC}$, $J_{SC}$, and PCE remain nearly constant only up to an interface defect density of approximately $10^{11}$ $cm^{-2}$, beyond which they decrease progressively until approximately $10^{15}$ $cm^{-2}$. Further increases in defect density produce little additional variation, indicating that the device enters a recombination-limited regime in which interface-mediated carrier losses have already become dominant. The stronger sensitivity of the rear interface suggests that efficient hole extraction at the CNGS/$MoTe_2$ junction is critically dependent on interface quality. Defect states at this interface provide recombination pathways for electrons reaching the back junction and holes transported through the HTL, thereby reducing carrier collection efficiency and suppressing quasi-Fermi level splitting. The observed saturation beyond $10^{15}$ $cm^{-2}$ indicates that the recombination activity has reached a level where further increases in trap density no longer significantly alter the dominant carrier-loss mechanism.

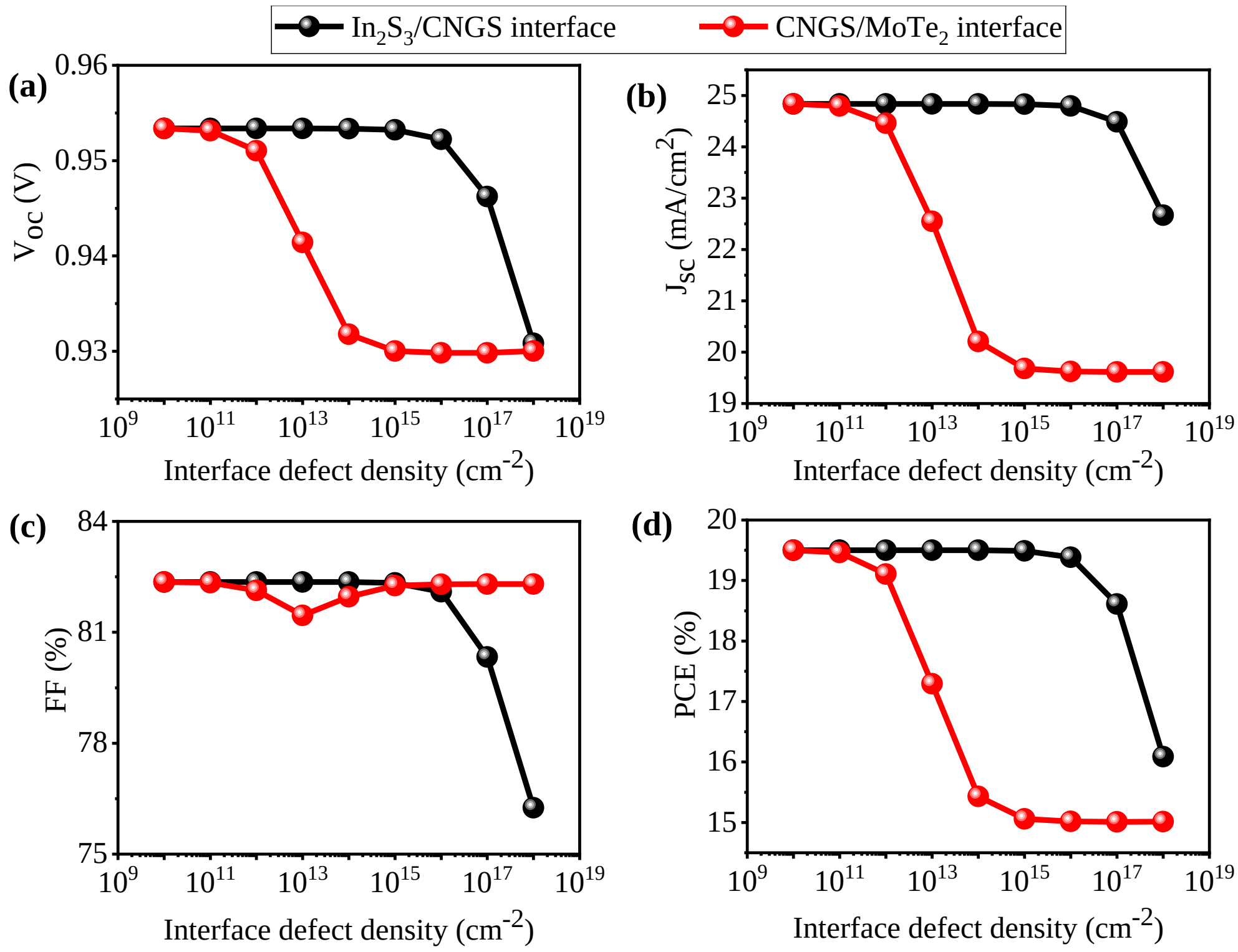

Fig. 7. Effect of defect density at charge transport layer (CTL) and absorber interfaces ($In_2S_3$/CNGS and CNGS/$MoTe_2$ interfaces) on (a) $V_{OC}$, (b) $J_{SC}$, (c) FF, and (d) PCE of the device. All other material parameters were kept fixed at their default values, as listed in Tables 1 and 2.

Unlike the other photovoltaic parameters, FF remains nearly unchanged throughout the investigated range of defect densities at the CNGS/$MoTe_2$ interface. This behavior suggests that the interface defects primarily affect carrier recombination rather than charge transport resistance. Consequently, the reduction in device performance is manifested predominantly through losses in voltage and photocurrent, while the shape of the current-voltage characteristics remains largely unaffected.

### 4.4. Series and shunt resistance

The electrical performance of practical solar cells is strongly influenced by resistive losses associated with series resistance ($R_{series}$) and shunt resistance ($R_{shunt}$) [72]. Series resistance originates from the finite conductivity of the absorber and transport layers, contact resistance at metal-semiconductor interfaces, and resistance within the electrodes [68], [73]. In contrast, shunt resistance is associated with parasitic leakage pathways arising from pinholes, grain boundaries, interfacial imperfections, and local short-circuit channels [73], [74]. Since both parameters directly affect carrier transport and power extraction, their optimization is essential for realizing high-performance photovoltaic devices.

The combined variation of $R_{series}$ and $R_{shunt}$ in this work reveals that shunt resistance exerts the dominant influence on device performance within the investigated resistance range (Fig. 8). $V_{OC}$

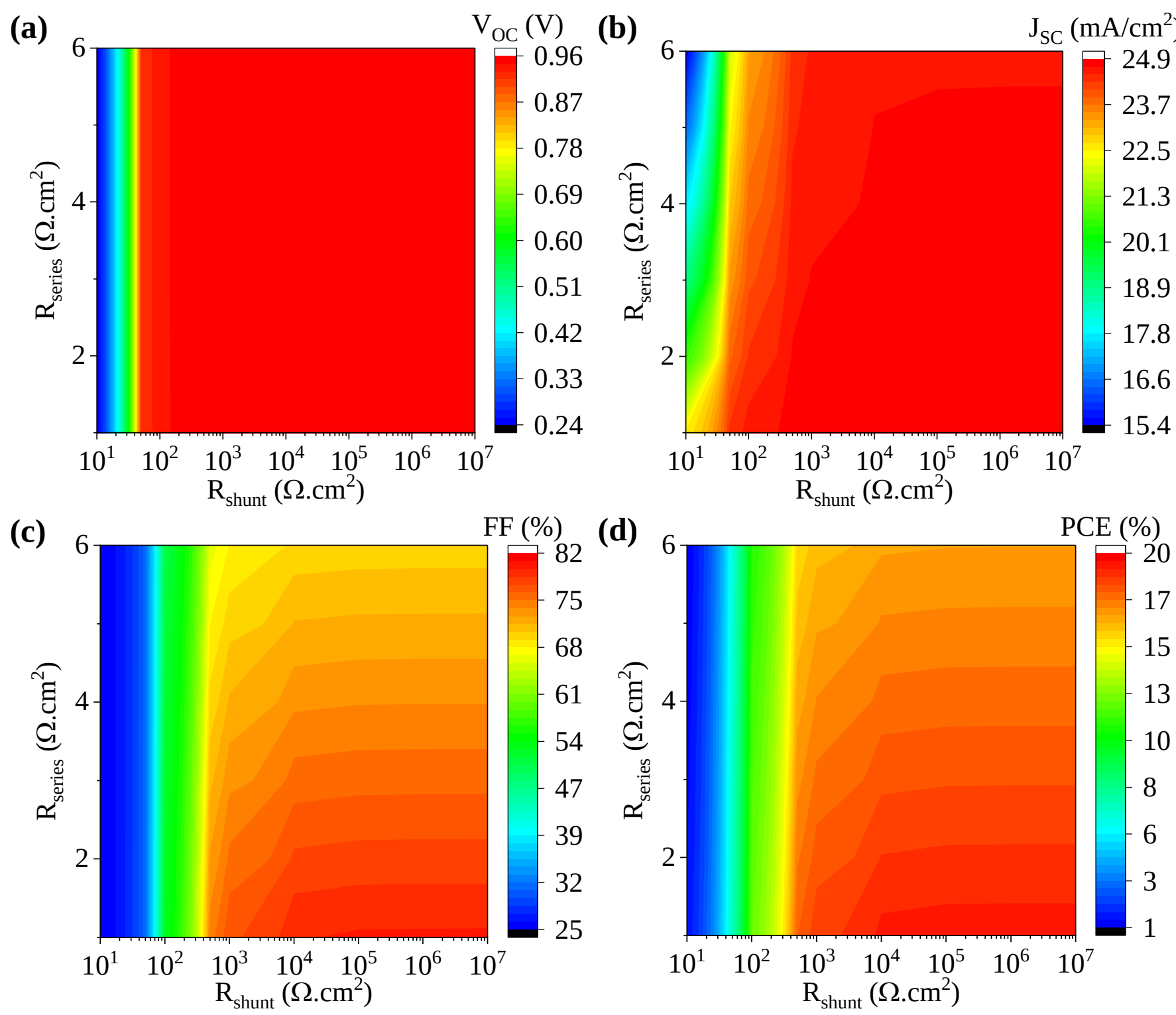

Fig. 8. Two-dimensional maps showing the simultaneous effects of shunt resistance ($R_{shunt}$) and series resistance ($R_{series}$) on the photovoltaic performance of the device: (a) $V_{OC}$, (b) $J_{SC}$, (c) FF, and (d) PCE. All other material and device parameters were kept fixed at their baseline values specified in Tables 1 and 2.

remains nearly constant for $R_{shunt}$ above approximately $10^2$ Ω.cm$^2$, indicating that leakage-current losses are effectively suppressed in this regime. When $R_{shunt}$ falls below this threshold, $V_{OC}$ decreases due to the emergence of parasitic current pathways that increase the dark current and reduce the attainable quasi-Fermi level splitting. In contrast, $V_{OC}$ remains essentially independent of $R_{series}$ throughout the investigated range because no net current flows under open-circuit conditions, eliminating any voltage drop across the series resistance.

$J_{SC}$, FF, and PCE exhibit a stronger dependence on the combined resistive losses. For $R_{shunt}$ exceeding approximately $10^3$ Ω·cm$^2$, these parameters remain nearly unchanged, suggesting that leakage-current losses have become negligible and that the investigated $R_{series}$ range (0-6 Ω·cm$^2$) does not introduce significant transport limitations. However, as $R_{shunt}$ decreases below $10^3$ Ω·cm$^2$, the influence of $R_{series}$ becomes increasingly apparent. Under these conditions, photogenerated carriers are partially diverted through leakage pathways while simultaneously experiencing transport losses associated with finite series resistance. The combined effect reduces carrier collection efficiency, distorts the current-voltage characteristics, and lowers the maximum extractable power, resulting in progressive degradation of $J_{SC}$, FF, and PCE.

The pronounced sensitivity to low shunt resistance highlights the importance of suppressing leakage-current pathways through high-quality film deposition and interface engineering. Meanwhile, maintaining low series resistance remains essential for minimizing transport losses and preserving efficient charge extraction. The results therefore demonstrate that optimal device performance requires both high $R_{shunt}$ and low $R_{series}$, ensuring minimal parasitic leakage and efficient carrier transport throughout the device structure.

### 4.5. Effect of radiative and auger recombination

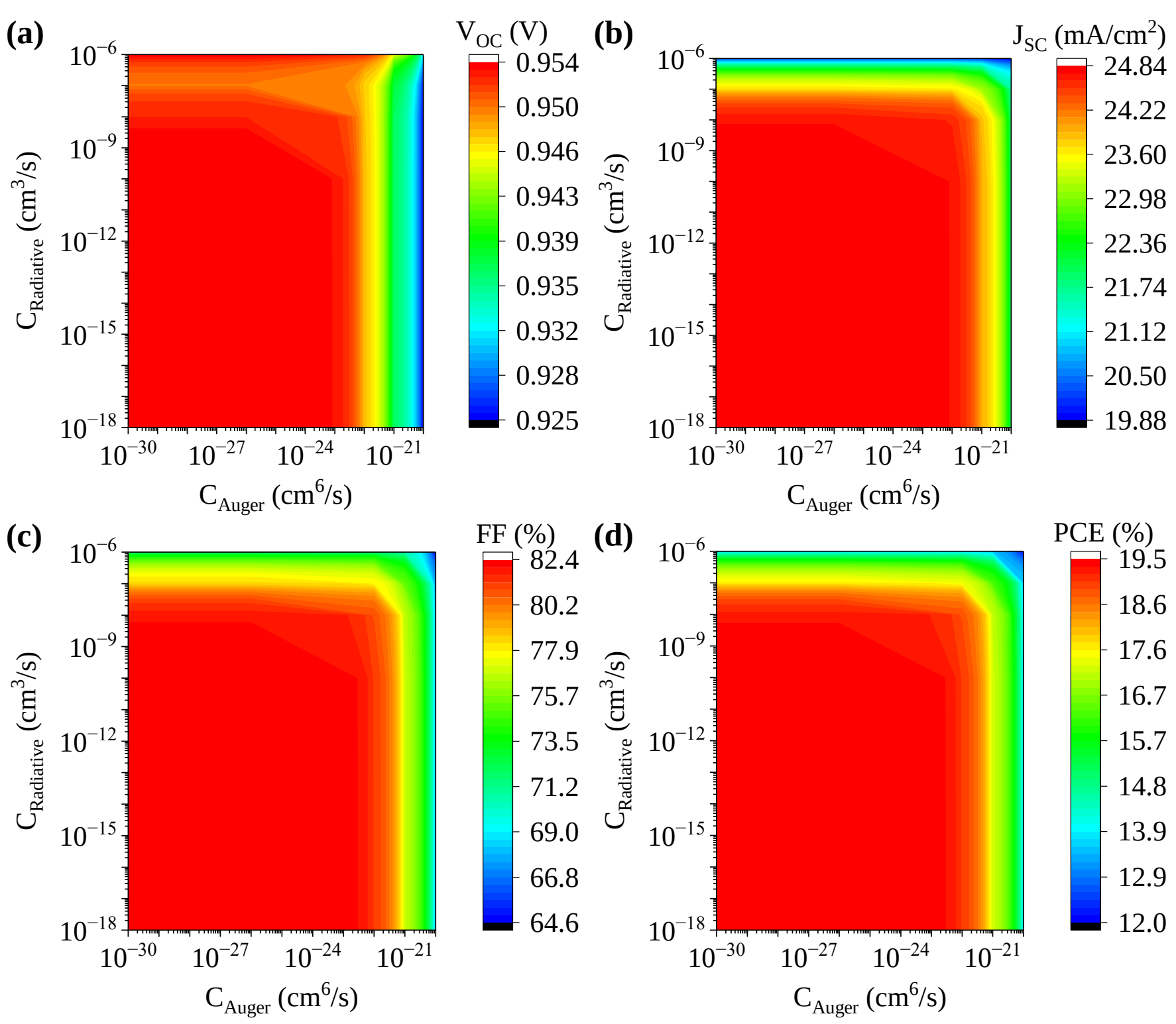

Fig. 9. Two-dimensional maps illustrating the coupled influence of Auger and radiative recombination coefficients on (a) $V_{OC}$, (b) $J_{SC}$, (c) FF, and (d) PCE of the device. All other material and device parameters were maintained at their baseline values specified in Tables 1 and 2.

To further elucidate the influence of intrinsic recombination processes on device performance, the radiative and Auger recombination coefficients were systematically varied while keeping all other material and device parameters unchanged (Fig. 9). These analyses provide insight into the recombination tolerance of the proposed device and identify the onset of performance degradation associated with increased carrier-loss mechanisms.

The simulated photovoltaic parameters exhibit excellent stability over a wide range of recombination coefficients. For radiative recombination, $V_{OC}$, $J_{SC}$, FF, and PCE remain nearly unchanged until the radiative recombination coefficient approaches approximately $10^{-7}$ $cm^3s^{-1}$. Beyond this threshold, all performance parameters gradually deteriorate. A similar trend is observed for Auger recombination, although the degradation begins at a substantially smaller coefficient of approximately $10^{-21}$ $cm^6s^{-1}$. These results indicate that the proposed device is relatively insensitive to moderate variations in intrinsic recombination coefficients but becomes increasingly susceptible once the corresponding recombination rates become comparable to the carrier extraction rate.

The observed behavior can be understood from the carrier recombination kinetics. Radiative recombination arises from the direct band-to-band annihilation of electrons and holes, with a recombination rate proportional to the product of the electron and hole concentrations (Equation 10). At relatively small values of the radiative recombination coefficient, this process constitutes only a minor fraction of the total carrier-loss mechanism, allowing photogenerated carriers to be efficiently separated and collected before recombination occurs. However, as the radiative recombination coefficient increases, the probability of direct electron-hole recombination rises, reducing the excess carrier population available for current generation. Consequently, the enhanced recombination lowers the quasi-Fermi level splitting, diminishes carrier collection efficiency, and results in simultaneous reductions in $V_{OC}$, $J_{SC}$, FF, and PCE.

Auger recombination exhibits a similar threshold-dependent behavior but is inherently more sensitive to carrier concentration because it involves a three-particle interaction in which the recombination energy is transferred to a third carrier rather than being emitted as a photon. Since the Auger recombination rate scales with carrier density (Equation 11), increasing the Auger coefficient rapidly accelerates non-radiative carrier losses once a critical value is exceeded. This additional recombination pathway shortens the effective carrier lifetime, increases the probability of carrier annihilation before extraction, and progressively degrades the photovoltaic performance.

Overall, the results demonstrate that the proposed solar cell exhibits strong tolerance to intrinsic radiative and Auger recombination over the physically relevant range of recombination coefficients. Significant performance degradation occurs only when the corresponding recombination coefficients exceed approximately $10^{-7}$ $cm^3s^{-1}$ for radiative recombination and $10^{-7}$ $cm^3s^{-1}$ for Auger recombination, highlighting the effectiveness of the proposed device architecture in maintaining efficient carrier extraction while suppressing intrinsic recombination losses.

### 4.6. Temperature-dependent performance analysis

Understanding temperature dependence is essential for assessing the operational stability of photovoltaic devices under realistic outdoor conditions. In this work, thermal behavior was investigated over the range of 275–500 K (Fig. 10). $V_{OC}$ decreases monotonically with increasing temperature. This behavior arises from the temperature dependence of the reverse saturation current density, which increases due to enhanced thermally activated carrier generation and increased SRH recombination. As temperature rises, the probability of carrier occupation of higher energy states increases, leading to stronger recombination activity and a reduction in quasi-Fermi level splitting. Consequently, $V_{OC}$ continuously degrades with temperature.

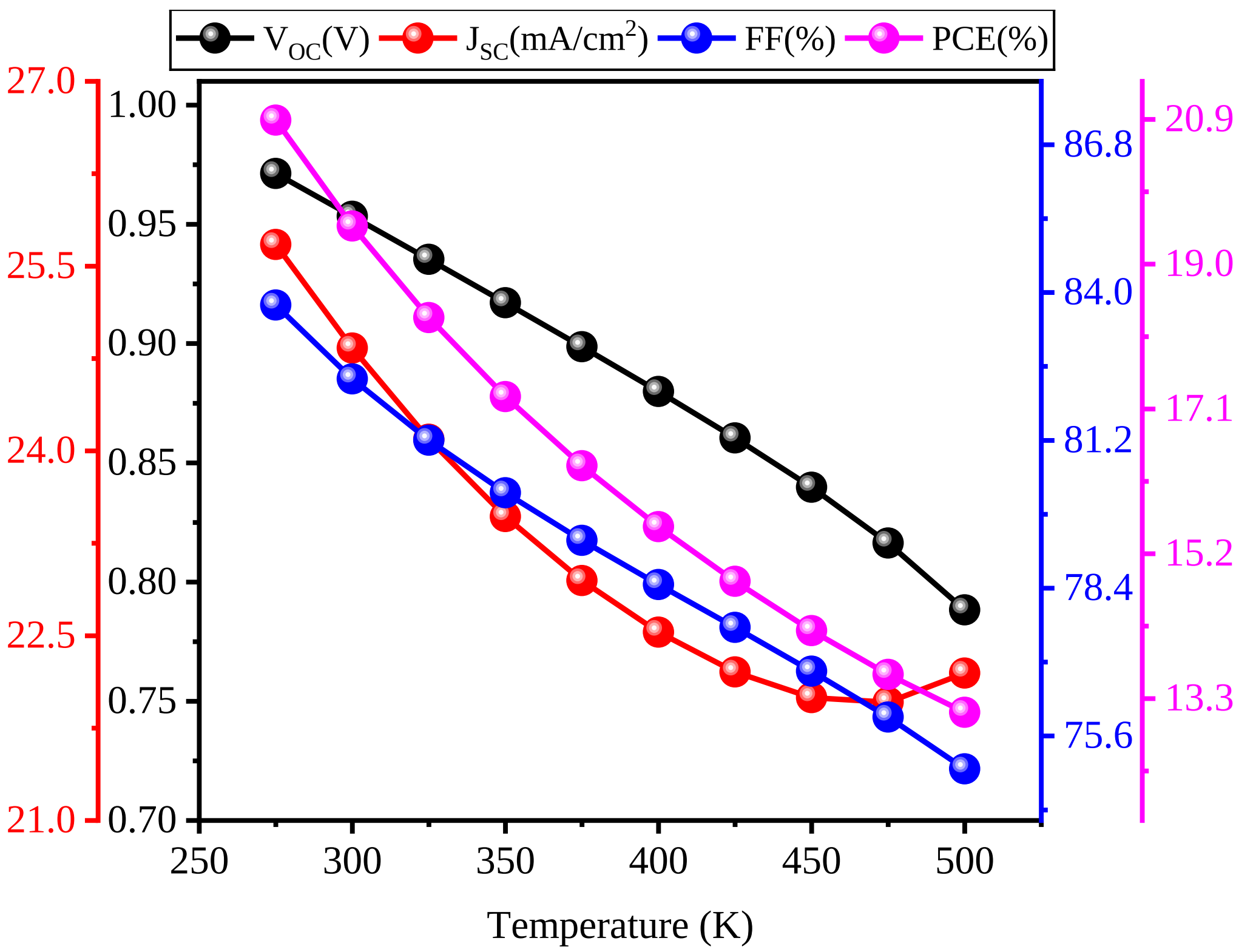

Fig. 10. Effect of temperature (T) on $V_{OC}$, $J_{SC}$, FF, and PCE of the device. All other material parameters were kept fixed at their default values, as listed in Tables 1 and 2.

FF also decreases steadily with increasing temperature. Elevated temperature increases carrier scattering and recombination rates, which reduce carrier lifetime and degrade transport efficiency. In addition, the weakening of the built-in electric field at higher temperatures reduces carrier separation efficiency, leading to increased resistive and recombination losses. These effects collectively result in a deterioration of the current–voltage characteristics and a reduction in FF.

$J_{SC}$ exhibits a weak non-monotonic dependence on temperature. In the lower temperature regime (275–450 K), $J_{SC}$ decreases gradually with increasing temperature due to enhanced recombination and reduced carrier diffusion length, which limit carrier collection efficiency. However, beyond approximately 450 K, a slight increase in $J_{SC}$ is observed. At higher temperatures, increased thermal energy enhances the emission rate of carriers from trap and defect states, reducing carrier trapping probability and partially improving carrier extraction efficiency. This thermally assisted detrapping effect counterbalances recombination losses, leading to a modest recovery in photocurrent at elevated temperatures.

PCE decreases continuously over the entire temperature range. Although a slight recovery in $J_{SC}$ is observed at higher temperatures, the pronounced reductions in $V_{OC}$ and FF dominate the device response. The increase in saturation current and recombination activity with temperature significantly reduces the overall energy conversion efficiency, resulting in a monotonic decline in PCE.

Overall, the results indicate that the proposed CNGS-based solar cell follows the expected thermal degradation behavior of heterojunction photovoltaic devices [75]–[77]. These findings highlight the importance of thermal stability considerations in device design, particularly for maintaining long-term operational efficiency under varying environmental conditions.

### 4.7. Illumination intensity dependence

The effect of light intensity on the photovoltaic performance of the device was investigated over the range of 100-1000 W/m$^2$ to evaluate the device response under varying illumination conditions (Fig. 11). Understanding light-intensity dependence is essential for assessing the operating behavior of solar cells under real outdoor environments, where irradiance fluctuates continuously due to atmospheric and seasonal variations.

$V_{OC}$ increases logarithmically with increasing light intensity. This behavior is consistent with the diode equation (13), where $V_{OC}$ depends on the logarithm of the photocurrent density relative to the saturation current [78]. As illumination increases, the photogenerated carrier density rises, leading to an increase in quasi-Fermi level splitting. However, because recombination processes also increase with carrier concentration, the growth of $V_{OC}$ follows a sub-linear, logarithmic trend rather than a linear one.

$J_{SC}$ exhibits an almost linear increase with light intensity across the entire investigated range. This linearity indicates that photogeneration scales proportionally with incident photon flux, while collection efficiency remains relatively stable under varying illumination.

FF shows a non-monotonic dependence on light intensity. At low to moderate illumination levels (up to approximately 300 W/m$^2$), FF decreases due to increased recombination activity and enhanced carrier accumulation effects within the device, which distort the current-voltage characteristics. As light intensity increases further, FF begins to recover and increases gradually. This improvement can be attributed to stronger photogenerated carrier extraction and the dominance of drift-assisted transport under higher injection conditions, which helps to reduce the relative impact of series resistance and improves the overall diode quality factor. PCE increases logarithmically with light intensity throughout the investigated range, resulting from the combined effects of $V_{OC}$, $J_{SC}$ and FF.

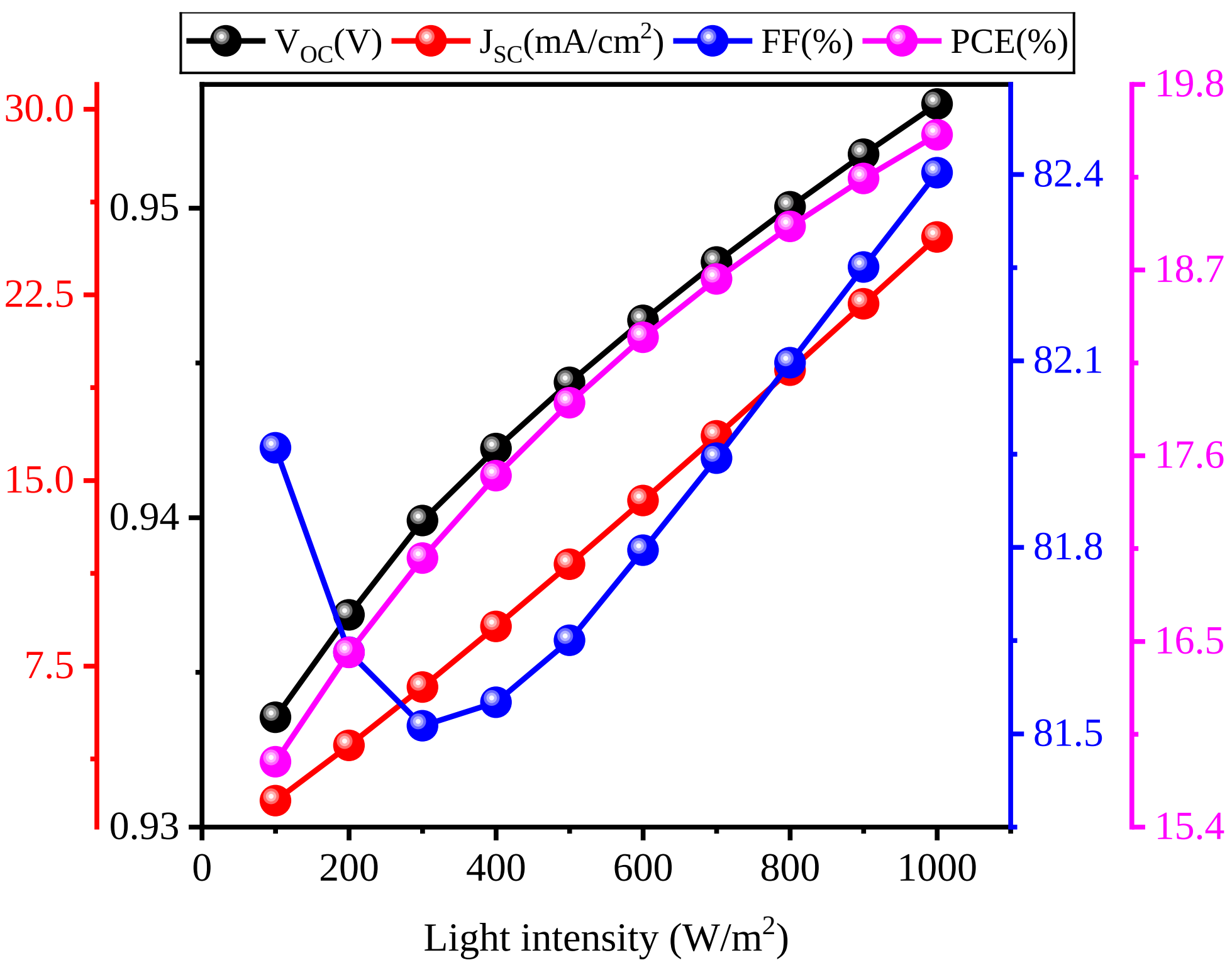

Fig. 11. Effect of incident light intensity on $V_{OC}$, $J_{SC}$, FF, and PCE of the device. All other material parameters were kept fixed at their default values, as listed in Tables 1 and 2.

Table 3. Comparison of photovoltaic performance between the proposed structure and literature-reported solar cell designs with CNGS as the absorber

| Device structure | $V_{OC}$ (V) | $J_{SC}$ (mA/cm$^2$) | FF (%) | PCE (%) | Reference | Year |
|---|---|---|---|---|---|---|
| AZO/iZnO/CdS/CNGS/Mo | 0.444 | 28.635 | 49.17 | 6.25 | [18] | 2022 |
| AZO/iZnO/CdS/CNGS/$MoS_2$/Mo | 0.648 | 29.039 | 58.07 | 10.94 | [18] | 2022 |
| AZO/iZnO/$Cd_{0.8}Zn_{0.2}S$/CNGS/$MoS_2$/Mo | 0.983 | 29.67 | 66.77 | 20.05 | [17] | 2023 |
| ZnO:Al/ZnO/$ZrS_2$/CNGS/SnS/Mo | 1.1009 | 22.89 | 83.99 | 21.17 | [79] | 2026 |
| **Al/AZO/$In_2S_3$/CNGS/$MoTe_2$/Au** | **0.984** | **34.39** | **84.04** | **28.44** | **This work** | **2026** |

## 4.8 Optimized device

Based on the systematic analyses presented in the preceding sections, the key material and device parameters were jointly optimized to obtain the highest photovoltaic performance of the proposed Al/AZO/$In_2S_3$/CNGS/$MoTe_2$/Au architecture under standard operating conditions of 300 K and 1000 $Wm^{-2}$ AM1.5G illumination. The optimized device achieves a $V_{OC}$ of 0.984 V, $J_{SC}$ of 34.39 mA/cm$^2$, FF of 84.04%, and PCE of 28.44%. To place the predicted performance in context, the optimized device was compared with previously reported CNGS-absorber-based solar cells, as summarized in Table 3. Reported PCEs for CNGS-based devices range from 6.25% to 21.17%, whereas the optimized architecture proposed in this work achieves 28.44%, exceeding the highest reported value in the literature, found till now. This improvement can be attributed to the synergistic optimization of the CNGS absorber and the $In_2S_3$/$MoTe_2$ charge-selective layers, together with favorable junction electrostatics, suppressed bulk and interfacial recombination, and extended spectral utilization through the $MoTe_2$ secondary-absorber contribution. Nevertheless, the predicted efficiency should be regarded as a theoretical performance projection under the adopted simulation framework and optimized material parameters, rather than as a direct indication of immediately achievable experimental efficiency. The comparison instead highlights the substantial photovoltaic potential of CNGS and provides experimentally relevant guidance for the development and optimization of CNGS-based thin-film solar cells.

The favorable simulated performance and the compatibility of the constituent materials with established thin-film fabrication techniques motivate further consideration of experimental realization. The potential fabrication pathways and experimental feasibility of the proposed architecture are discussed in the following section.

## 5. Conclusion

In this work, a comprehensive numerical investigation of a CNGS-based thin-film solar cell with the Al/AZO/$In_2S_3$/CNGS/$MoTe_2$/Au architecture was presented, with particular emphasis on the interplay between carrier transport, recombination, and junction electrostatics. The optical analysis demonstrated that $MoTe_2$ can contribute to long-wavelength photon absorption while simultaneously functioning as the HTL, thereby providing additional spectral utilization. Systematic optimization of the absorber and charge-selective layers demonstrated that device performance is governed by the balance between optical absorption, electrostatic field strength, carrier transport, and recombination losses**.** The defect analysis revealed markedly different interface tolerances, with the $In_2S_3$/CNGS junction being considerably more tolerant than the CNGS/$MoTe_2$ interface, highlighting the latter as a critical target for interface engineering. The resistance analysis further confirmed the importance of maintaining high shunt resistance and low series resistance. Temperature and illumination studies showed the expected thermal degradation of photovoltaic performance and enhanced device response under stronger illumination, while the C-V, C-f, and Mott-Schottky analyses provided further insight into depletion behavior, junction electrostatics, and photoinduced charge dynamics. Following comprehensive optimization, the proposed device achieves $V_{OC}$ = 0.984V, $J_{SC}$ = 34.39 mA/cm$^2$, FF = 84.04%, and PCE = 28.44% under AM1.5G illumination at 300 K. This predicted efficiency exceeds the previously reported 6.25–21.17% range for CNGS-based solar cells considered in this study, highlighting the potential of the $In_2S_3$/$MoTe_2$ charge-selective-layer combination. Importantly, this value represents a theoretical performance projection within the assumptions of the SCAPS-1D model and should not be interpreted as an experimentally demonstrated efficiency. Collectively, these findings establish $Cu_2NiGeS_4$ as a promising yet comparatively underexplored absorber for thin-film photovoltaics and provide physically motivated guidelines for its further development. In particular, controlling absorber quality, minimizing interfacial defects especially at the CNGS/$MoTe_2$ junction and reducing parasitic resistive losses are identified as key requirements for approaching the predicted performance experimentally. Given the compatibility of the constituent materials with established thin-film fabrication approaches, the proposed architecture offers a promising platform for experimental investigation and further development of high-efficiency CNGS-based photovoltaics.

## Acknowledgements

The authors gratefully acknowledge Dr. M. Burgelman of the University of Gent in Belgium for kindly providing the SCAPS-1D software.

## Author Contributions

Md Tashfiq Bin Kashem was involved in the conceptualization, SCAPS-1D methodology, investigation, data analysis, visualization, supervision, and in writing and revising the manuscript. Hasib Md Abid Bin Farid was involved in the SCAPS-1D methodology, data analysis, visualization, and in revising the manuscript.

## Funding

The authors declare that no funds, grants, or other support were received during the preparation of this manuscript.

## Declaration of Competing Interest

The authors declare that they have no known competing financial interests or personal relationships that could have appeared to influence the work reported in this paper.

## Data Availability

The data supporting the findings of this study are available from the corresponding author upon reasonable request.